\documentclass{iopjournal}

\usepackage{graphicx}
\usepackage{amsmath,amssymb,bm,booktabs,array,multirow,makecell}

\usepackage{multirow}
\usepackage{xcolor}
\usepackage{caption}
\usepackage{floatrow}
\usepackage{makecell}
\newcommand{\AHS}{A_{\mathrm{HS}}}

\newcommand{\Eexact}{E_{\mathrm{exact}}}

\fancyhfoffset[L]{0pt}
\renewcommand{\headrulewidth}{0pt}
\renewcommand{\footrulewidth}{0pt}

\begin{document}


\title{Symmetry-guided algorithms for variational quantum optimization}

\author{Eunjin In$^{1,*}$, Kristina D. Launey$^1$, Darin C. Mumma$^{1}$ and James A. Sauls$^2$}


\affil{$^1$Department of Physics and Astronomy, Louisiana State University, Baton Rouge, Louisiana 70803, USA}

\affil{$^2$Hearne Institute of Theoretical Physics, Louisiana State University, Baton Rouge, Louisiana 70803, USA}

\affil{$^*$Author to whom any correspondence should be addressed.}

\email{ein1@lsu.edu}


\begin{abstract}
Quantum numbers associated with symmetries classify quantum states and identify the sectors relevant to a given simulation. Generic hardware-efficient ansatzes do not necessarily preserve the quantum numbers and may therefore explore states outside the target sector. 
We develop symmetry-guided algorithms based on the quantum-accessible Hilbert--Schmidt asymmetry (\(\AHS\)), which quantifies the deviation of a quantum state from invariance under a specified symmetry group.  
This quantity guides the optimization toward a highly reduced symmetry-based target sector, without enforcing the symmetry on the preparation circuit. 
We demonstrate the approach as a proof of principle in variational quantum eigensolver calculations targeting singlet ground states with collective SU(2) symmetry.
The applications include a five-qubit \(S\)-wave \(p+{}^3\mathrm{He}\) two-cluster Hamiltonian, deduced from \textit{ab initio} many-body calculations, and collective spin models with up to 12 particles, which provide a many-particle benchmark of asymmetry-guided optimization.
Across these applications, \(\AHS\)-based guidance increases the singlet probability. 
Overall, the proposed algorithms improve energy accuracy during optimization and ground-state fidelity compared with direct energy minimization, with substantial fidelity improvements in the larger spin systems.
These results demonstrate the utility of an asymmetry objective determined by the group action, independently of the Hamiltonian energy, for improving the reliability of variational quantum optimization.

\end{abstract}

\section{Introduction}
\label{Introduction}

Preparing suitable input states is an essential step in quantum simulation. 
Variational methods prepare states by optimizing parameterized quantum circuits against a chosen objective \cite{McClean2016Theory,Cerezo2021VQA,Bharti2022NISQ}, but the ability of a circuit to represent a desired state does not guarantee that the optimization will 
converge to that state \cite{Cerezo2021VQA,McClean2018,Larocca2025BarrenPlateaus}. 
Symmetry provides physical information that can guide this search. When a Hamiltonian conserves a symmetry, its Hilbert space separates into sectors characterized by the corresponding quantum numbers.  
General hardware-efficient ansatzes  
do not necessarily preserve these quantum numbers \cite{2020Gard,Ryabinkin2019Constrained,2021Kuroiwa}, motivating the use of an explicit symmetry criterion to guide state preparation toward the target sector.

Several strategies incorporate symmetry into variational calculations.
Symmetry-based encodings reduce the qubit register needed to represent a selected sector \cite{Bravyi2017Tapering}, including approaches exploiting point-group and other Boolean symmetries \cite{Picozzi2023SAE}.
Related work in quantum machine learning develops systematic methods for incorporating symmetry into parametrized quantum circuits \cite{Meyer2023Equivariant,Nguyen2024Equivariant}.
Symmetry-preserving state preparation circuits confine the search to a chosen sector when their input state belongs to that sector \cite{2020Gard,2023Lyu}. For instance,  
Ref.~\cite{Crognaletti2025Equivariant}
exploits total-spin and translational symmetries to resolve singlet and triplet levels in a frustrated spin chain. 
However, imposing additional symmetries on the circuit can introduce implementation costs. 
In the noisy molecular calculations of Ref.~\cite{2020Gard}, circuits 
enforcing additional spin symmetries require more CNOT gates, which can reduce the accuracy of the computed energies. 
Other approaches incorporate symmetry information through penalty terms in the variational objective \cite{Ryabinkin2019Constrained,2021Kuroiwa} or projection onto a selected sector \cite{Lacroix2020SymmetryPreparation,2023Stetcu,2024Rule}. 
Symmetry verification uses measurement outcomes to mitigate contributions inconsistent with the desired quantum numbers \cite{2018Bonet,McArdle2019SymmetryVerification}.

A warm start (WS) uses information from  a preparatory calculation to initialize a variational algorithm \cite{Egger2021WarmStarting,Puig2025WarmStarts}. 
Input circuit parameters can be obtained by pretraining with classically optimized matrix product states \cite{Dborin2022Pretraining}, using approximate amplitude encoding to obtain circuit parameters from an approximate state \cite{Truger2025AmplitudeWarmStart}, or transferring parameters between related Hamiltonians or system sizes \cite{Skogh2023ParameterTransfer,Gibbs2025NuclearSymmetry}. 
Input states for subsequent calculations can also be prepared by selecting components according to energy or symmetry.
The rodeo algorithm prepares Hamiltonian eigenstates through energy filtering using controlled Hamiltonian evolution and postselection on auxiliary-qubit measurements \cite{2021Choi,CohenOh2023Rodeo}.
Symmetry projection can prepare states with specified quantum numbers \cite{2023Stetcu,2024Rule}, and is used in Ref.~\cite{Mihalikova2025} to prepare inputs for symmetry-preserving variational optimization. 
Another approach is to optimize the circuit parameters against a symmetry objective in a separate preparation stage \cite{Selvarajan2022symmetry}. 
Such a preparation need only establish the required symmetry content. It need not approximate the full target wave function.

The variational quantum eigensolver (VQE) provides a setting in which to assess 
the usefulness of such preparation. 
It combines quantum measurements with classical optimization to minimize the Hamiltonian expectation value and can use relatively shallow circuits, making it a promising approach for noisy intermediate scale quantum (NISQ) devices \cite{Cerezo2021VQA,Peruzzo2014,Kandala2017,Preskill2018NISQ}.
Its accuracy depends on both the variational flexibility of the ansatz and the effectiveness of parameter optimization \cite{Wecker2015Variational}.
ADAPT-VQE uses energy gradients to construct compact ansatzes and can achieve higher accuracy with shallower circuits than unitary coupled cluster with singles and doubles (UCCSD) \cite{Grimsley2019ADAPT}. 
With an unconstrained circuit, energy minimization can adjust both the symmetry sector probabilities and the correlations within them. 
Our approach uses symmetry information to concentrate the search on the target sector within a prescribed ansatz. Holding the prepared parameters fixed can further reduce the parameter space explored during energy minimization.

Ref.~\cite{2023Bandyopadhyay} developed efficient quantum algorithms 
for symmetry testing using measures based on the Hilbert--Schmidt distance.
For quantum states, purity and overlap measurements on two copies of a state make asymmetry directly accessible on a quantum device.
We refer to this measure as Hilbert--Schmidt asymmetry and denote it by \(\AHS\). It quantifies the departure of a density operator from invariance under a specified symmetry group and vanishes for invariant states.

In this work, we use \(\AHS\) as a variational objective to prepare states with the desired symmetry using generic circuits.
The preparation objective is defined directly by the group action, without requiring an explicit construction of Casimir operators or the amplitudes and preparation parameters of a known symmetric state.
We optimize the circuit parameters to minimize \(\AHS\) and then use the resulting parameters as a warm start for VQE, denoting this algorithm by $\mathrm{WS}_{\mathrm{symm}}$--VQE.
A symmetry preserving circuit maintains the prepared sector probabilities during energy minimization, 
whereas a generic circuit permits further changes. 
We therefore also introduce Symm--VQE, which repeatedly uses asymmetry information during the energy search by alternating parameter updates guided by $\AHS$ with energy minimization.
Beyond VQE, the preparation may also provide input states for other quantum algorithms requiring the corresponding symmetry.

We present two proof-of-principle applications
under collective SU(2) rotations,
with the spin models providing a many-particle benchmark.
For pure states, zero asymmetry identifies the singlet sector.
The nuclear application is motivated by the use of symmetry to organize many-body model spaces in \textit{ab initio} nuclear calculations \cite{DytrychLDRWRBB20,Launey2021} and of angular momentum to distinguish reaction channels in cluster descriptions of nuclei, including weakly bound systems \cite{Quaglioni2009,Viviani2011,Mercenne2022,launey2026_SAreview}. 
We consider a five qubit $S$-wave $p+{}^3$He two-cluster model with an effective non-local potential, deduced within an \textit{ab initio} many-body framework and represented in a finite configuration basis, to test symmetry guidance in circuits with angular and radial degrees of freedom. 
Collective spin models with up to 12 spins test preparation in a singlet sector containing multiple independent states, followed by exchange optimization that preserves the prepared sector probability. 
We assess changes in singlet probability, energy convergence, and ground-state fidelity. We also test whether preparation using finite samples of collective rotations reaches the prescribed asymmetry threshold across different initializations in the spin models.
We examine the effect of measurement shot noise on optimization in the nuclear model.

\section{Asymmetry-guided variational algorithms}
\label{sec:asymm-guided-var-algorithms}

\subsection{Symmetry objective and target states}
\label{sec:symm-obj-targ-states}

For a state $\rho$ and a unitary representation $\{U(g)\}$ of a compact group $G$,
we write the Hilbert--Schmidt asymmetry as
\begin{equation}
\AHS(\rho)=\int_G dg\,\|[U(g),\rho]\|_2^2,
\label{eq:ahs}
\end{equation}
where $dg$ is the normalized Haar measure and the Hilbert–Schmidt norm of an operator $X$ is $\|X\|_2^2=\operatorname{Tr}(X^\dagger X)$. This is the compact group form of the finite group measure in Ref.~\cite{2023Bandyopadhyay}.
With the group twirl 
$\mathcal{T}_G(\rho)=\int dg\,U(g)\rho U^\dagger(g)$,
it becomes
\begin{equation}
\AHS(\rho)=2\left[\operatorname{Tr}(\rho^2)-\operatorname{Tr}\!\left(\rho \mathcal{T}_G(\rho) \right)\right].
\label{eq:ahs-overlap}
\end{equation}
It vanishes precisely when $\rho$ is invariant under every group element. 
Such invariant states are the free states in the resource theory of asymmetry \cite{Gour2008ReferenceFrames}.
The purity and overlap terms can be estimated using Bell measurements on pairs of corresponding qubits in two copies of a state, as in the destructive SWAP test \cite{2023Bandyopadhyay,GarciaEscartin2013Swap}. For the overlap term, the group transformation is applied to one copy. 
Figure~\ref{fig:nuclear_and_spin_ansatzes}a in Appendix~\ref{app:circuits_ansatzes} illustrates this measurement circuit for collective SU(2) rotations.
Sampling $T$ independent group elements from the normalized Haar measure gives
\begin{equation}
\AHS^{\mathrm{sampled}}(\rho)=\frac{2}{T}\sum_{t=1}^{T}\left[\operatorname{Tr}(\rho^2)-\operatorname{Tr}\!\left(\rho U(g_t)\rho U^\dagger(g_t)\right)\right].
\label{eq:sampled-ahs}
\end{equation}
The superscript ``sampled" refers to the group average.

For $N$ spin-$1/2$ degrees of freedom, a collective SU(2) rotation acts through the same single-spin rotation on every spin,
\begin{align}
U(g) &= \bigotimes^{N-1}_{i=0} D^{(1/2)}(g)_{q_i},
\label{eq:general-collective-su2-rotation}\\
D^{(1/2)}(g) &= R_z(\alpha)R_y(\beta)R_z(\gamma).
\label{eq:spin-half-euler-rotation}
\end{align}
The group average of $U(g)\rho U^\dagger(g)$ can be sampled by drawing $\alpha$ and $\gamma$ independently and uniformly from $[0,2\pi)$ and $\cos\beta$ uniformly from $[-1,1]$. The evaluation method and number of sampled rotations are specified for each calculation.

On a quantum processor, $\AHS^{\mathrm{sampled}}$ in Eq.~\eqref{eq:sampled-ahs} can be estimated using only purity and rotated overlap measurements, as illustrated in Fig.~\ref{fig:nuclear_and_spin_ansatzes}a. 
To understand the underlying mechanism of the symmetry objective, we note that for a pure state, 
evaluating the SU(2) group average in Eq.~\eqref{eq:ahs-overlap} gives
\begin{equation}
\AHS(\rho) = 2\left[1-\sum_J\frac{\operatorname{Tr}(\rho_J^2)}{2J+1}\right].
\label{eq:ahs-general}
\end{equation}
The SU(2) average removes coherences between different $J$ sectors and completely depolarizes each irreducible representation space, while preserving the state on the multiplicity space \cite{2007Bartlett}.
The representation space has dimension $2J+1$, corresponding to $M=-J,\ldots,J$, giving the normalized identity $I_{2J+1}/(2J+1)$.
Here $\rho_J$ is a positive operator on the multiplicity space, which distinguishes independent copies of the spin-$J$ representation. Its trace, $\operatorname{Tr}(\rho_J)=P_J$, is the probability of the total spin sector $J$. 
Thus, in general, $\AHS$ depends on the operators $\rho_J$ as well as the sector probabilities.

We consider 
pure target states with total angular momentum $J=0$, whereas the method is applicable to any $J$ values. A pure state density operator is invariant under these rotations if and only if the state lies in the singlet sector, including any multiplicity of that sector. For mixed states, rotational invariance does not guarantee that the state is restricted to
the singlet sector. Minimizing \(\AHS\) over pure variational states therefore provides an objective for seeking and finding states confined to the singlet sector.
When this sector contains multiple independent states, an additional criterion is needed to select a particular state within it.

\subsection{Application to VQE: initialization and alternating optimization}
\label{sec:apptoVQE}

We use \(\AHS\) to guide optimization toward states with the desired symmetry content. In VQE, the parameters are also optimized to minimize the Hamiltonian expectation value,
\begin{equation}
E(\bm\theta) = \langle\psi(\bm\theta)|H|\psi(\bm\theta)\rangle,
\label{eq:energy-cost}
\end{equation}
where $|\psi(\bm\theta)\rangle$ is the parametrized trial state. We refer to the optimization of circuit parameters against $E(\bm\theta)$ as an energy minimization stage.

We combine asymmetry guidance and energy minimization in two ways: 
(1) A warm start algorithm, ${\mathrm{WS}_{\mathrm{symm}}}$--VQE, first optimizes the circuit parameters to minimize \(\AHS\). 
This step requires neither a classically computed approximation to the target wave function nor Hamiltonian energy evaluations.
The prepared state then provides the starting point for energy minimization.
A symmetry-preserving circuit retains the sector probabilities established during preparation, whereas a generic circuit can change them and may benefit from renewed asymmetry guidance. 
(2) An alternating algorithm, Symm--VQE, utilizes the asymmetry guidance during the VQE evolution, namely, it repeatedly uses asymmetry information to guide parameter updates and follows these updates with energy minimization. The final parameters of each stage initialize the next. 

We examine these possibilities in nuclear and spin models. The nuclear benchmark uses ${\mathrm{WS}_{\mathrm{symm}}}$--VQE and two alternating variants, Symm--VQE and Symm--VQE (unmixed), with generic circuits.
In the spin benchmark, ${\mathrm{WS}_{\mathrm{symm}}}$--VQE combines asymmetry-based preparation with subsequent energy optimization using SU(2)-preserving exchange gates, while holding the preparation parameters fixed.
The following sections specify the objectives, updated parameter blocks and evaluation costs for each implementation.

\section{Nuclear model benchmark}

\subsection{Hamiltonian and qubit encoding}

We follow the two-cluster Hamiltonian encoding
of Ref.~\cite{Soorya2026}. For the neutron-nucleus elastic-scattering
application discussed in that work, an effective nonlocal interaction $V_{N}^{(J)}$ is derived using the symmetry-adapted no-core shell model with the Green's function approach~\cite{launey2026_SAreview,burrows2024abpotentials,MummaBL2025}. This approach uses $(A\pm1)$-body eigenstates and energies, together with single-particle overlaps connecting the $A$-body target ground state to these eigenstates through neutron addition or removal. 
This construction connects microscopic nuclear structure to a two-cluster description.
Here we consider the $S$-wave $p+{}^3\mathrm{He}$ system, including the Coulomb interaction between the charged clusters. The proton and the ${}^3\mathrm{He}$ ground state both have spin $1/2$, so their coupling at relative orbital angular momentum $\ell=0$ gives the positive-parity ${}^{1}S_{0}$ $(J=0)$ and ${}^{3}S_{1}$ $(J=1)$ channels.
After separating the center-of-mass motion, the Hamiltonian for the relative motion at distance $r$ ($r'$) between the proton and target before (after) scattering in each channel is
\begin{equation}
H^{(J)}=\frac{p^2}{2\mu}
+V_{N}^{(J)}(r,r')
+V_C(r),
\label{eq:relative-hamiltonian}
\end{equation}
where $\mu=m_pm_{{}^3\mathrm{He}}/(m_p+m_{{}^3\mathrm{He}})$ is the reduced mass, $V_N^{(J)}$ is the channel-dependent nonlocal nuclear interaction, and $V_C$ is the Coulomb interaction. 
We represent the relative Hamiltonian in a spherical harmonic-oscillator basis characterized by $\hbar\Omega=20$ MeV, where $\Omega$ is the oscillator angular frequency. 
The radial basis is truncated to $n_r^{\mathrm{max}}=8$ states, labeled by $n_r=0,\ldots,7$, in each channel, yielding a real symmetric $8\times8$ matrix $H^{(J)}$. 
The lowest eigenstate of the truncated $J=0$ Hamiltonian has an energy of $-1.421$~MeV and serves as the target state for the variational calculation.

We retain the full two-spin space to assess asymmetry guidance with generic hardware-efficient ansatzes that need not preserve $J$ or $M$. 
Both the ${}^{1}S_{0}$ and ${}^{3}S_{1}$ channels are included, with $M=0$ for ${}^{1}S_{0}$ and $M=-1,0,1$ for ${}^{3}S_{1}$.
The channel construction uses the coupled basis $|(I_1,I_2)I,n_r,\ell;JM\rangle$, where $I_1=I_2=1/2$ are the cluster spins, $I$ is their coupled spin, and $J$ results from coupling $I$ to $\ell$. For the present $S$-wave channels, $\ell=0$ and $J=I$. 
Rotational invariance makes the Hamiltonian block diagonal in $J$ and $M$, with the $H^{(J)}$ for all $M$ at a given $J$. The basis is ordered by decreasing $J$ and $M$, with the radial index varying fastest.
For the variational calculation, we express the Hamiltonian in the uncoupled basis $|m_1,m_2,n_r\rangle$ using Clebsch-Gordan coefficients. Here $m_1$ and $m_2$ are the individual cluster spin projections. The orbital state $\ell=m_\ell=0$ is omitted from the basis labels and requires no qubit.

Qubits $q_0$ and $q_1$ encode $m_1$ and $m_2$, respectively, with $|0\rangle$ corresponding to $+1/2$ and $|1\rangle$ to $-1/2$. The remaining three qubits encode $n_r=0,\ldots,7$ in binary as $|q_4q_3q_2\rangle=|000\rangle,\ldots,|111\rangle$. The full five-qubit basis follows Qiskit's $|q_4q_3q_2q_1q_0\rangle$ ordering. For example, $|00110\rangle$ represents $|m_1=+1/2,m_2=-1/2,n_r=1\rangle$.
Energy evaluations use the full five-qubit states, whereas $\AHS$ estimation uses two copies of the two-qubit spin state in the measurement circuit of Fig.~\ref{fig:nuclear_and_spin_ansatzes}a.

\subsection{Reference state and ansatzes}

The reference state is the product $|\phi_{\mathrm{ref}}\rangle=|m_1=\tfrac12,m_2=-\tfrac12,n_r=0\rangle$
with equal probabilities in the $J=0$ and $J=1$ sectors.
We prepare the trial state by applying a hardware-efficient circuit to this reference. 
Figure~\ref{fig:nuclear_and_spin_ansatzes}b in Appendix~\ref{app:circuits_ansatzes} shows the two-layer circuit with the two spin qubits above the three radial qubits.
Each variational layer applies an $R_y$ rotation followed by an $R_z$ rotation to every qubit, followed by independently parameterized $R_{xx}$ gates on $(q_0,q_1)$, $(q_2,q_3)$, and $(q_3,q_4)$. A final layer of $R_y$ and $R_z$ rotations completes the circuit. The two-qubit gates can entangle the two spins or the radial qubits, but no gate couples the spin and radial degrees of freedom. 
The two-layers circuit used here has 36 variational parameters.
Appendix~\ref{app:nuclear-entanglement} extends the comparison to a circuit that allows entanglement between the spin and radial degrees of freedom.

\subsection{Spin objective and optimization algorithms}

\begin{figure}[htbp]
\centering
\includegraphics[width=0.95\textwidth]{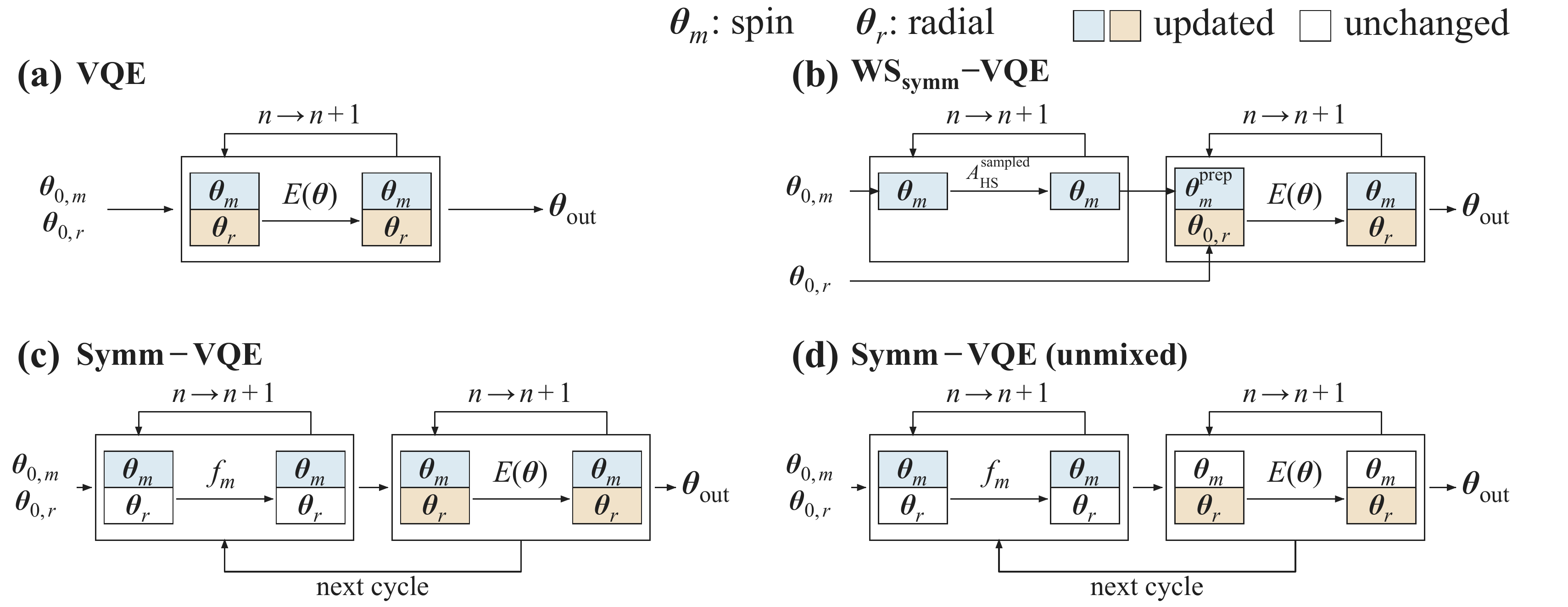}
\caption{Optimization algorithms for the nuclear model. 
Blue and beige boxes indicate updated spin and radial parameters, respectively, while white boxes indicate unchanged parameters.
Upper loops denote optimizer updates, $n\rightarrow n+1$, within each stage, and lower loops denote alternating cycles.
(a) VQE minimizes $E(\bm\theta)$ over both the spin and radial parameters, $\bm\theta_m$ and $\bm\theta_r$.
(b) ${\mathrm{WS}_{\mathrm{symm}}}$--VQE first minimizes $\AHS^{\mathrm{sampled}}$ over $\bm\theta_m$. The resulting $\bm\theta_m^{\mathrm{prep}}$ and the initial radial parameters $\bm\theta_{0,r}$ initialize energy minimization over both parameters.
(c) Symm-VQE alternates spin optimization using the conditional objective $f_m$ in Eq.~\eqref{eq:angular-stage-objective} with energy minimization over both parameters.
(d) Symm-VQE (unmixed) uses the same spin stage but updates only $\bm\theta_r$ during the energy stage. 
Here $\bm\theta_{\mathrm{out}}$ denotes the final parameters. 
Each stage ends when COBYLA terminates or its budget is reached. WS preparation allows 50 asymmetry evaluations. Alternating spin and energy stages allow 50 and 200 optimizer iterations, respectively. Alternation ends after four cycles or when consecutive cycle end energies differ by at most $10^{-3}$ MeV. VQE and ${\mathrm{WS}_{\mathrm{symm}}}$--VQE allow at most 1,000 total optimizer iterations, including preparation for $\AHS$.
}
\label{fig:algorithm-schedules}
\end{figure}

Figure~\ref{fig:algorithm-schedules} illustrates the optimization procedure for the four algorithms, while Table~\ref{tab:protocols} summarizes the objectives and parameters varied in each stage.
The variational parameters are partitioned as
$\bm{\theta}=\left(\bm{\theta}_{m},\bm{\theta}_{r}\right),$
where $\bm{\theta}_{m}$ and $\bm{\theta}_{r}$ parametrize the spin projection and radial circuits, respectively. 
The trial state factorizes as
$|\psi(\bm\theta)\rangle=|\psi_{\rm spin}(\bm\theta_m)\rangle \otimes |\psi_{\rm radial}(\bm\theta_r)\rangle$.
The collective rotations defined in Eqs.~\eqref{eq:general-collective-su2-rotation} and~\eqref{eq:spin-half-euler-rotation} apply the same spin-$1/2$ rotation to both cluster spins and act as the identity on the radial degrees of freedom. Since the trial state factorizes into spin and radial components, its asymmetry equals that of the spin state.
We evaluate $\AHS^{\mathrm{sampled}}$ in Eq.~\eqref{eq:sampled-ahs} on the two-qubit spin state, while the energy objective $E(\bm\theta)$ in Eq.~\eqref{eq:energy-cost} is evaluated on the full five-qubit state and depends on both parameter blocks.

Direct energy minimization, denoted by VQE, varies both parameter blocks throughout the search and provides the baseline comparison.
The warm-start algorithm,
${\mathrm{WS}_{\mathrm{symm}}}$--VQE, first minimizes $\AHS^{\mathrm{sampled}}$ over $\bm\theta_m$, holding $\bm\theta_r$ at its initial value. 
The resulting spin parameters and the unchanged radial parameters then initialize energy minimization over both blocks.

\begin{table}[htbp]
\caption{Objectives and parameters varied in each stage of the four optimization algorithms.
A single arrow denotes preparation followed by energy minimization, while a double arrow denotes alternating optimization, beginning with the spin stage. Only the listed parameters are varied in each stage, with the remaining parameters fixed. Here $\bm{\theta}=\left(\bm{\theta}_{m},\bm{\theta}_{r}\right)$. 
See also Fig.~\ref{fig:algorithm-schedules}.
}
\label{tab:protocols}
\centering
\scriptsize
\setlength{\tabcolsep}{3pt}
\begin{tabular}{@{}lcc@{}}
\toprule
Algorithm & Objective sequence & Varied parameters\\
\midrule
VQE & $E(\bm\theta)$ & $\bm{\theta}$\\
\addlinespace
WS$_{\rm Symm}$-VQE & $\AHS^{\mathrm{sampled}}\;\longrightarrow\;E(\bm\theta)$ & $\bm{\theta}_{m}\;\longrightarrow\;\bm{\theta}$\\
\addlinespace
Symm-VQE & $f_m\;\longleftrightarrow\;E(\bm\theta)$ & $\bm{\theta}_{m}\;\longleftrightarrow\;\bm{\theta}$\\
\addlinespace
Symm-VQE (unmixed) & $f_m\;\longleftrightarrow\;E(\bm\theta)$ & $\bm{\theta}_{m}\;\longleftrightarrow\;\bm{\theta}_{\mathrm{r}}$\\
\bottomrule
\end{tabular}
\end{table}

Since energy minimization with a generic circuit can change the prepared symmetry content, the alternating algorithms, Symm-VQE and Symm-VQE (unmixed),
revisit the spin parameters as the radial state is optimized.
Each cycle begins with a spin stage, in which $\bm\theta_m$ is varied while $\bm\theta_r$ is held at its current value. This is the initial radial parameter vector in the first cycle and the result of the preceding energy stage in subsequent cycles.
The sampled asymmetry determines whether a trial state is assessed by its energy or by an asymmetry-dependent cost,
\begin{equation}
f_{m}(\bm{\theta_m}; \bm{\theta_r})=
\begin{cases}
E(\bm\theta_m, \bm\theta_r), & \AHS^{\mathrm{sampled}}(\bm{\theta_m})\leq\tau,\\[3pt]
\displaystyle
\frac{|E_{\mathrm{ref}}|}
{\sqrt{1-\AHS^{\mathrm{sampled}}(\bm{\theta_m})/2}},
& \AHS^{\mathrm{sampled}}(\bm{\theta_m})>\tau,
\end{cases}
\label{eq:angular-stage-objective}
\end{equation}
where the threshold is set empirically to $\tau=0.8$ and $\AHS^{\mathrm{sampled}}$ is the finite rotation estimate in Eq.~\eqref{eq:sampled-ahs} evaluated on the spin circuit.
The choice of the asymmetry-dependent cost is discussed in Appendix~\ref{app:conditional-objective}.
The reference energy $E_{\mathrm{ref}}$ is  the energy reached at the end of the preceding energy stage. 
For the first spin stage, it is evaluated at the initial parameter vector. It remains fixed throughout each spin stage.
For nonzero $E_{\rm ref}$, the asymmetry-dependent cost 
increases with $\AHS^{\mathrm{sampled}}$ for $\tau<\AHS^{\mathrm{sampled}}<2$. Minimizing this cost therefore favors lower asymmetry without evaluating the energy of each trial state. 
When $\AHS^{\mathrm{sampled}}\leq \tau$, the spin parameters are instead optimized using $E(\bm\theta)$, with the radial parameters held fixed. 
This empirical switching rule uses $|E_{\mathrm{ref}}|$ to set the cost scale in the adopted energy convention.
The subsequent energy stage minimizes $E(\bm\theta)$ throughout, without applying the switching rule.
Symm-VQE varies both $\bm\theta_m$ and $\bm\theta_r$, allowing the spin and radial states to adjust together. Symm-VQE (unmixed) fixes the spin parameters and varies only $\bm\theta_r$, preserving the spin state and its sector probabilities during this stage. 
Both algorithms repeat the stages, with the parameters reached at the end of each stage initializing the next.

For the statevector calculations,
the sampled asymmetry in Eq.~\eqref{eq:sampled-ahs} is evaluated using a set of $T=24$ Haar random rotations shared across all nuclear calculations. The sample size is chosen using the Hoeffding bound in Ref.~\cite{2023Bandyopadhyay}, $T=2\epsilon^{-2} \rm \ln(2/\delta)$, with an additive error $\epsilon=0.5$ and a failure probability $\delta=0.1$ for estimating the group average of a given state.

\subsection{Simulation results}
\label{sec:nuclear-simulation-results}

\begin{figure}[htbp]
\centering
\includegraphics[width=\textwidth]{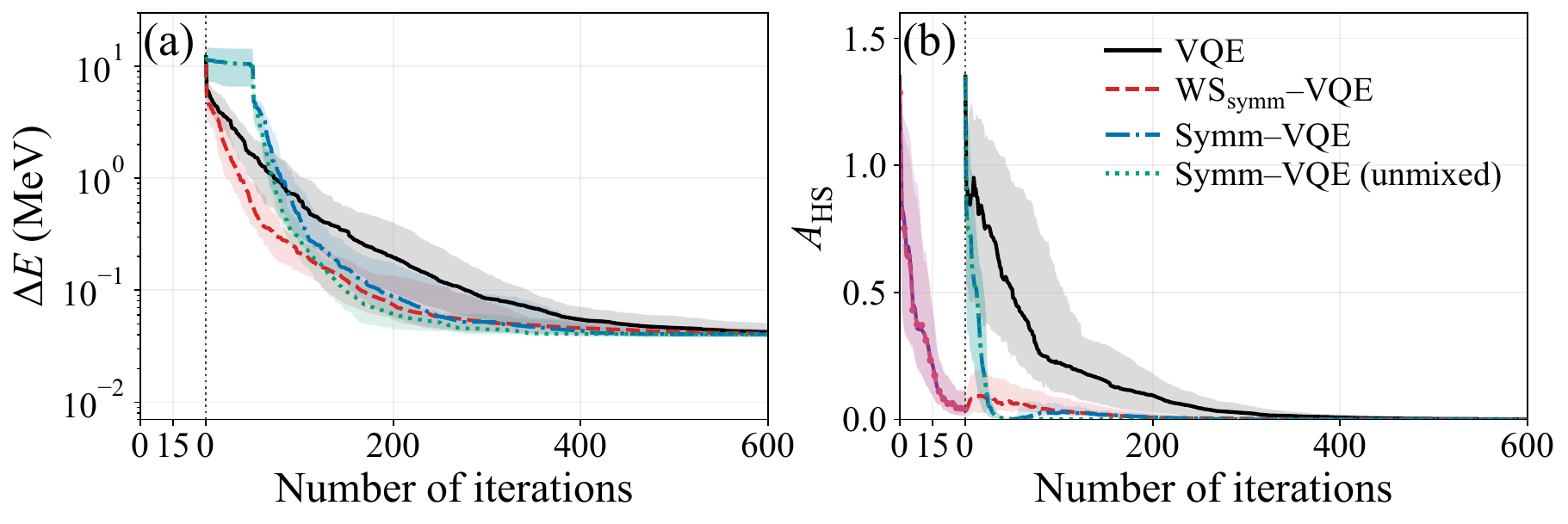}
\caption{Convergence of the three symmetry-guided algorithms based on $\AHS$ compared to the conventional VQE: (a) $\Delta E=E-\Eexact$, where $\Eexact$ is the lowest eigenvalue of the truncated $p+{}^3{\rm He}$ Hamiltonian, and (b) exact Haar averaged asymmetry from Eq.~\eqref{eq:ahs-general}. 
Curves and shaded bands show medians and interquartile ranges over 50 initial parameter vectors drawn from the $0.2\,\mathcal{N}(0,1)$ normal distribution. 
All calculations use the two-layer circuit and COBYLA with statevector evaluations.
Preparation for ${\mathrm{WS}_{\mathrm{symm}}}$--VQE is shown to the left of the vertical dotted line in (b). Preparation takes 19--30 iterations (median 24), with final values carried forward to iteration 30 for runs ending earlier. 
At the dotted line, the ${\mathrm{WS}_{\mathrm{symm}}}$--VQE curve shows the prepared state and the other algorithms show their initial states.
}
\label{fig:noiseless-iteration-trajectories}
\end{figure}

The nuclear benchmark examines how asymmetry guidance affects convergence and the variation in the resulting states across initializations. Figure~\ref{fig:noiseless-iteration-trajectories} shows the energy error and exact asymmetry for the three $\AHS$-guided algorithms and VQE, obtained from statevector simulations.
In panel (a), the three $\AHS$-guided algorithms generally yield smaller median errors at the same iterations and show quicker convergence to comparable energy accuracy than direct VQE.
The alternating algorithms initially devote iterations to spin optimization, after which their energy errors decrease to values comparable to those of ${\mathrm{WS}_{\mathrm{symm}}}$--VQE. 

In addition, as shown in Fig.~\ref{fig:noiseless-iteration-trajectories}b, the guided algorithms 
approach rotational invariance earlier than the direct VQE. For ${\mathrm{WS}_{\mathrm{symm}}}$--VQE, the initial reduction in asymmetry occurs during preparation before the energy objective is introduced. 
The reduction in $\AHS$ is accompanied by an increase in the singlet probability $P(J=0)$, evaluated from the simulated statevector
by summing the squared amplitudes of the $J=0$ components over all radial basis states in the coupled basis. This probability is used only for analysis and does not enter the optimization.
Its median increases from 0.49 to 0.99 during preparation, with interquartile range (IQR) of $[0.47, 0.50]$ and $[0.97, 1.00]$, respectively. These ranges describe the variation across initializations. The increase occurs in all 50 runs, showing that the preparation consistently increases support in the target sector for the initial states considered.
For the alternating algorithms, the use of the asymmetry-dependent cost is concentrated in the first spin stage. In the first cycle, the median percentage of spin stage objective function calls for which $\AHS^{\mathrm{sampled}} > \tau=0.80$ is $23.03\%$ for both algorithms. This percentage decreases to approximately $3\%$ in subsequent cycles. 
All four algorithms achieve median fidelities $F > 
99.5\%$, where $F=|\langle\psi_0|\psi\rangle|^2$ and $|\psi_0\rangle$ is the lowest energy eigenstate obtained by exact diagonalization of the model Hamiltonian.

To compare measurement requirements, 
we estimate $N_{\rm{evaluations}}=144N_E + 25N_A$, where $N_E$ and $N_A$ count energy and asymmetry evaluations across all stages, including preparation. The coefficient 144 counts the Pauli terms in the Hamiltonian expansion, including the identity term with known expectation. Each $\AHS^{\mathrm{sampled}}$ evaluation comprises one purity term and 24 rotated overlap terms, for a total of 25 terms. ${\mathrm{WS}_{\mathrm{symm}}}$--VQE, Symm--VQE and Symm--VQE (unmixed) reduce the median estimated count required to reach the joint energy--asymmetry accuracy as $\Delta E < 0.5$ MeV and $\AHS<0.05$ by $57.6\%$, $48.9\%$ and $60.5\%$, respectively, relative to VQE. The reduced number of energy evaluations offsets the additional asymmetry evaluations in these comparisons. This estimate does not account for Pauli grouping and shots, and thus does not directly quantify the total cost on quantum hardware.

\begin{figure}[htbp]
\centering
\includegraphics[width=0.85\textwidth]{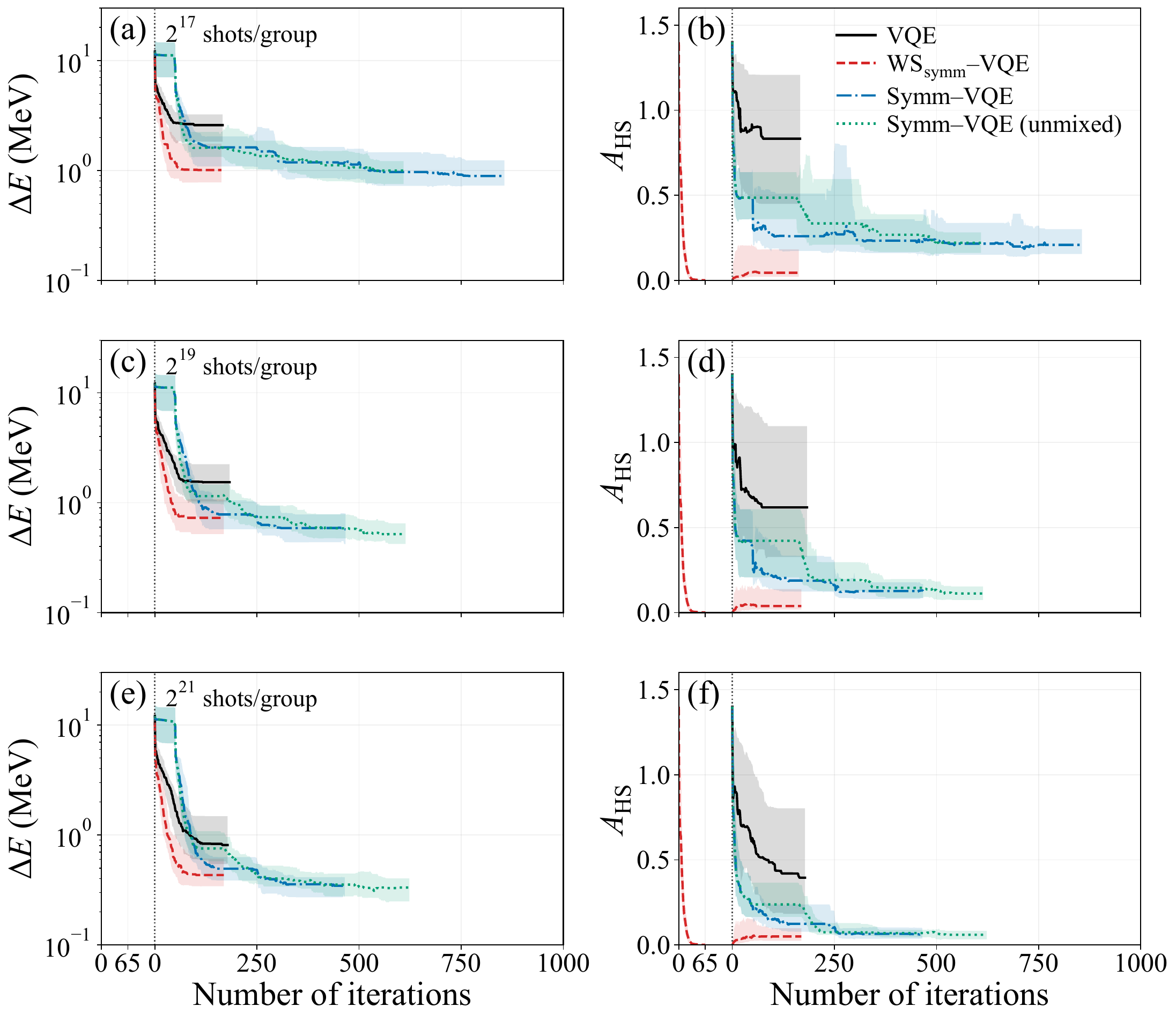}
\caption{
Energy error (left) and sampled asymmetry (right) for the four algorithms under finite-shot optimization. Rows correspond to $2^{17},2^{19},2^{21}$ energy shots per qubit-wise commuting group. The plotted values are evaluated from statevectors at each iteration of the finite-shot optimization. Asymmetry uses the same $T=96$ rotations as the finite-shot objective. Colors and line styles follow Fig.~\ref{fig:noiseless-iteration-trajectories}. Curves and shaded bands show medians and interquartile ranges over the same 50 initial parameter vectors. Preparation for ${\mathrm{WS}_{\mathrm{symm}}}$--VQE is shown to the left of the vertical dotted line.
}
\label{fig:nuclear-finite-shots-SV-analysis}
\end{figure}

We also examine how measurement shot noise affects the four algorithms.
Figure~\ref{fig:nuclear-finite-shots-SV-analysis} shows the energy error and sampled asymmetry during finite-shot optimization. 
The calculations use the same two-layer circuit and 50 initial parameter vectors as in Fig.~\ref{fig:noiseless-iteration-trajectories}. 
Asymmetry measurements use $T=96$ rotations and $4,096$ shots for each purity and rotated-overlap measurement. Energy measurements use $2^{17},2^{19}$, or $2^{21}$ shots for each of the 42 qubit-wise commuting groups.
Only measurement shot noise is included, providing an estimate of performance in the far-term error-corrected regime. Energy, asymmetry, and ground state fidelity are evaluated from statevectors at each iteration of the finite-shot optimization. 
Appendix~\ref{app:finite-shot-nuclear-raw-data} shows the corresponding measurement estimates.

At $2^{17}$ energy shots per group, the median final energy errors are $2.582$, $1.009$, $0.890$, and $0.998$ MeV for VQE, ${\mathrm{WS}_{\mathrm{symm}}}$--VQE, Symm--VQE and Symm--VQE (unmixed), respectively. The corresponding median fidelities are $70.78\%$, $95.23\%$, $92.14\%$, and $91.97\%$. 
At $2^{19}$ shots, the median energy errors decrease to 1.537, 0.728, 0.442, and 0.519 MeV, while the median fidelities increase to $80.71\%$, $95.85\%$, $96.21\%$, and $95.75\%$. 
At $2^{21}$ shots, these errors decrease further to 0.811, 0.433, 0.329, and 0.333 MeV, with median fidelities of $88.48\%$, $97.52\%$, $98.08\%$, and $97.56\%$
Increasing the energy shots thus reduces the median energy error and increases the median fidelity for all four algorithms.
At each shot count, the three $\AHS$-guided algorithms attain smaller median energy errors and higher median fidelities than VQE.
Although increasing the energy shots reduces the VQE energy error, its median asymmetry remains nonzero even at $2^{21}$ shots per group. The improved energy accuracy is therefore not accompanied by full recovery of the target symmetry.
In the right panels, the asymmetry decreases during ${\mathrm{WS}_{\mathrm{symm}}}$--VQE preparation but increases during subsequent energy minimization, when both spin and radial parameters are varied. The alternating algorithms reduce the asymmetry further during later spin stages.

\section{Extension to collective spin models}
\label{sec:spin-models}

\subsection{Model and variational circuit}

The nuclear model includes both the spins of the two clusters and their relative radial motion.
We next consider larger systems containing only spins, for which the spin space itself contains multiple independent singlets.  
In this application, energy minimization uses an exchange circuit that preserves probabilities of the total spin sectors. This allows us to examine whether preparing the symmetry content provides useful inputs for ground-state calculations as the number of spins (particles) increases.

We consider a collective spin model motivated by neutrino interactions~\cite{Neill2024}, described by a Heisenberg-like Hamiltonian with all-to-all couplings. This class of Hamiltonian has been used in VQE calculations~\cite{Mihalikova2025} and in quantum simulations of collective neutrino oscillations~\cite{Hall2021,Amitrano2023}. 
The Hamiltonian considered here takes the form
\begin{equation}
H_\nu=\sum_{i<j}g_{ij}\,\bm\sigma_i\cdot\bm\sigma_j,
\qquad g_{ij}=1-\widehat{\bm p}_i\cdot\widehat{\bm p}_j.
\label{eq:neutrino-hamiltonian}
\end{equation}
Each pair interaction is invariant under collective SU(2) rotations, so the Hamiltonian preserves total angular momentum. The momentum directions $\widehat{\bm p}_i$ are sampled isotropically by drawing $\cos\vartheta_i$ uniformly from $[-1,1]$ and $\varphi_i$ uniformly from $[0,2\pi)$. One Hamiltonian is used at each size $N=4,8,10,12$, and exact diagonalization confirms a singlet ground state in each case. The $N=6$ instance has a $J=1$ ground state and 
is considered in Appendix~\ref{app:N6_spin_cases} as an example of a non-singlet target.

The variational state is constructed by applying a preparation circuit followed by an exchange circuit to the reference $|\Phi_p\rangle$,
\begin{equation}
|\psi(\bm\alpha,\bm\beta)\rangle
=U_{\mathrm{ex}}(\bm\beta)U_{\mathrm{prep}}(\bm\alpha)|\Phi_p\rangle,
\label{eq:spin-trial}
\end{equation}
where $\bm\alpha$ and $\bm\beta$ denote the preparation and exchange parameters, respectively. The reference $|\Phi_p\rangle$ aligns the Bloch vector of spin $i$ with $\widehat{\bm p}_i$, following the momentum-aligned construction discussed in Refs.~\cite{Mihalikova2025,Neill2024}.
Figure~\ref{fig:nuclear_and_spin_ansatzes}c in Appendix~\ref{app:circuits_ansatzes} illustrates the preparation and exchange circuits with an example of four particles.
The preparation circuit consists of an initial layer of local rotations, disjoint CX gates on pairs (0,1), (2,3), $\cdots$, and a final layer of local rotations. Each local rotation layer applies $R_y$ followed by $R_z$ to every qubit, and each CX gate uses the lower-index qubit as the control. The circuit has $4N$ independent parameters and $N/2$ CX gates. 
We use one layer of exchange gates over all spin pairs. 
On pair $(i,j)$, the gates $R_{xx}(\beta_{ij})$, $R_{yy}(\beta_{ij})$ and $R_{zz}(\beta_{ij})$ share an angle and implement
\begin{equation}
U_{ij}(\beta_{ij})=
\exp\!\left[-\frac{i\beta_{ij}}{2} \left(X_iX_j+Y_iY_j+Z_iZ_j\right)\right].
\label{eq:spin-exchange-gate}
\end{equation}
There are $N(N-1)/2$ independent exchange parameters. The exchange gate order is randomized once and held fixed across runs at each size.
Each gate commutes with collective SU(2) rotations, preserving both the sector probabilities and the full rotational asymmetry. Energy optimization can therefore change the correlations within the prepared sectors while leaving their probabilities unchanged.

\subsection{Asymmetry state-preparation and exchange optimization}

We compare VQE with our warm-start procedure ${\mathrm{WS}_{\mathrm{symm}}}$--VQE  that now combines the $\AHS$ state-preparation with the exchange optimization.
VQE minimizes the Hamiltonian energy over all parameters $(\bm\alpha,\bm\beta)$.
${\mathrm{WS}_{\mathrm{symm}}}$--VQE first minimizes $\AHS$ over $\bm\alpha$, then holds these parameters fixed and minimizes the energy over the exchange parameters $\bm\beta$. Both methods use the same 50 initial parameter vectors at each size, with $\bm\alpha_0=0.1\bm\xi_\alpha$ and $\bm\beta_0=0.01\bm\xi_\beta$, respectively, where the components of each $\bm\xi$ vector are independent standard normal variables.
The momentum-aligned reference state is fixed at each system size.
The comparison thus examines the effect of preparing the input and subsequently restricting the energy search to exchange parameters.
For preparation, the exact asymmetry is evaluated using Eq.~\eqref{eq:ahs-general}.

\subsection{Simulation results}

\begin{figure}[tbp]
\centering
\includegraphics[width=.7\textwidth,height=.86\textheight,keepaspectratio]{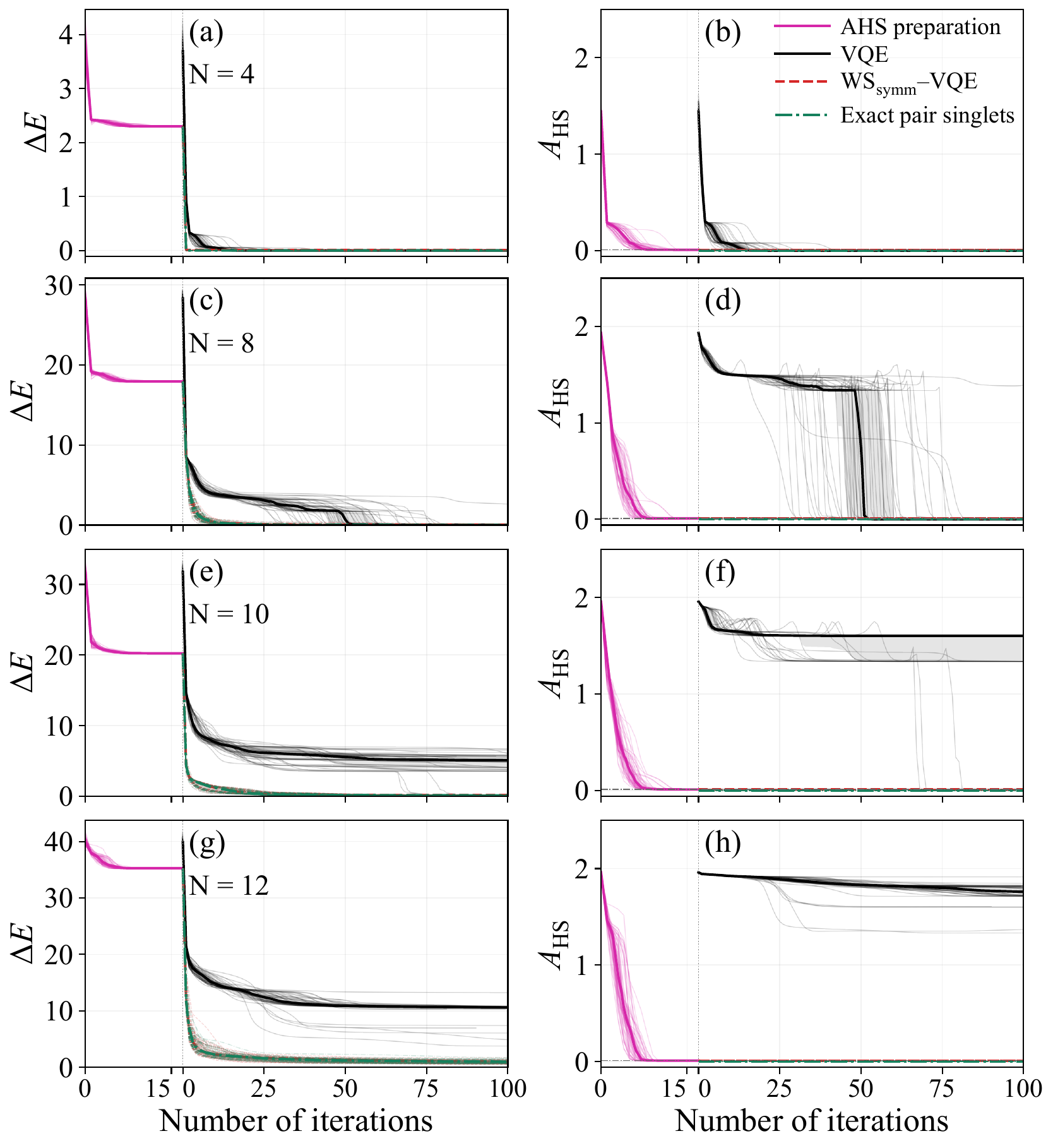}
\caption{Energy error and exact asymmetry $\AHS$ during preparation and exchange optimization for $N=4,8,10,12$ (rows). 
Magenta denotes $\AHS$ preparation, black VQE and red dashed ${\mathrm{WS}_{\mathrm{symm}}}$--VQE. 
Green dash-dotted curves show exchange optimization from exact pair singlets.
Thin curves show 50 initializations, and thick curves and bands give medians and interquartile ranges with terminal values carried forward. Preparation is shown to the left of the vertical dotted line, with ticks at 0 and 15 preparation sweeps. Iterations restart at zero for energy optimization. The energy shown during preparation is evaluated for comparison and is not used in the preparation objective.
All three methods use Rotosolve coordinate optimization \cite{Ostaszewski2021structure}. 
One iteration corresponds to a complete sweep over the parameters optimized in each stage. The calculations use statevector simulations. Preparation stops at $\AHS<10^{-2}$. 
}
\label{fig:spin-exact-ahs-trajectories}
\end{figure}

Figure~\ref{fig:spin-exact-ahs-trajectories} shows the energy error and exact $\AHS$ during preparation and the subsequent energy search. In the preparation interval, the asymmetry decreases to below $10^{-2}$ for all 50 initializations at each size. The corresponding median singlet probability increases from $46.93\%$, $3.94\%$, $3.97\%$ and $4.37\%$ for $N=4,8,10,12$, respectively, to above $99.75\%$ at each size. The energy need not decrease during this interval, since preparation minimizes asymmetry rather than energy. The prepared states provide the inputs for exchange optimization, which preserves their singlet probabilities within numerical precision.

At $N=4$, both methods rapidly reduce the energy error. At $N=8$, ${\mathrm{WS}_{\mathrm{symm}}}$--VQE approaches low energy in fewer energy stage sweeps, while VQE also converges as optimization proceeds. Both methods attain high final fidelities for these smaller systems. At $N=8$, the initial energy reduction in direct VQE precedes the substantial decrease in asymmetry, showing that lower energy can be obtained while appreciable weight remains outside the singlet sector. The difference is more pronounced at $N=10$ and 12. VQE retains substantial energy error and asymmetry for most initializations, whereas ${\mathrm{WS}_{\mathrm{symm}}}$--VQE lowers the energy while preserving the asymmetry below $10^{-2}$ established during preparation.
Ground-state fidelity confirms the improvement in state accuracy at the larger sizes. At the end of the full runs, the median fidelity at $N=10$ is $99.10\%$, with interquartile range of $98.91$--$99.19\%$, for ${\mathrm{WS}_{\mathrm{symm}}}$--VQE, compared with $29.63\%$ (IQR: $6.83$--$31.18\%$) for VQE. At $N=12$, the corresponding values are $95.27\%$ (IQR: $93.94$--$95.98\%$) and $0.02\%$ (IQR: $0.00$--$5.46\%$), respectively. 
For comparison, we use directly prepared products of two spin singlets with the same exchange circuit, initial exchange parameters, and optimization settings as ${\mathrm{WS}_{\mathrm{symm}}}$--VQE (green dash-dotted curves in Fig.~\ref{fig:spin-exact-ahs-trajectories}. For $N=4,8,10,12$, the median final energy errors are approximately zero, 0.0058, 0.0922, and 0.7713, respectively, compared with 0.0065, 0.0394, 0.1174, and 0.7919 following $\AHS$ preparation. The corresponding median fidelities are $100\%$, $99.97\%$, $99.34\%$, and $95.31\%$, compared with $99.77\%$, $99.75\%$, $99.10\%$, and $95.27\%$ following $\AHS$ preparation.
Thus, using only the group action, $\AHS$ preparation finds inputs that yield comparable final ground state fidelities after exchange optimization.
Appendix~\ref{app:spin-additional-entanglers} presents results with additional entangling gates between the spin pairs. These gates improve VQE, but its final fidelities remain substantially below those of ${\mathrm{WS}_{\mathrm{symm}}}$--VQE at $N=10$ and 12. 

Although ${\mathrm{WS}_{\mathrm{symm}}}$--VQE reaches a high final fidelity at $N=12$, its median ground-state fidelity decreases from approximately $0.07\%$ to $0.01\%$ during asymmetry preparation. By comparison, the spin projection procedure of Ref.~\cite{Mihalikova2025}
increases the fidelity from $2.4\%$ to $56.7\%$ before exchange optimization in their neutrino example.
Spin projection and $\AHS$-based preparation both establish the symmetry content before minimizing the energy, but they act differently within the singlet sector.
Spin projection extracts the singlet component of the initial state, whereas $\AHS$ minimization adjusts the preparation circuit without selecting a particular singlet. 
In the present calculation, $\AHS$ preparation concentrates the state in the singlet sector, whose probability is then preserved by the exchange circuit. Energy optimization therefore proceeds with the state remaining predominantly in the target subspace, even though preparation has not increased its ground-state overlap.

\begin{figure}[tbp]
\centering
\includegraphics[width=0.95\textwidth]{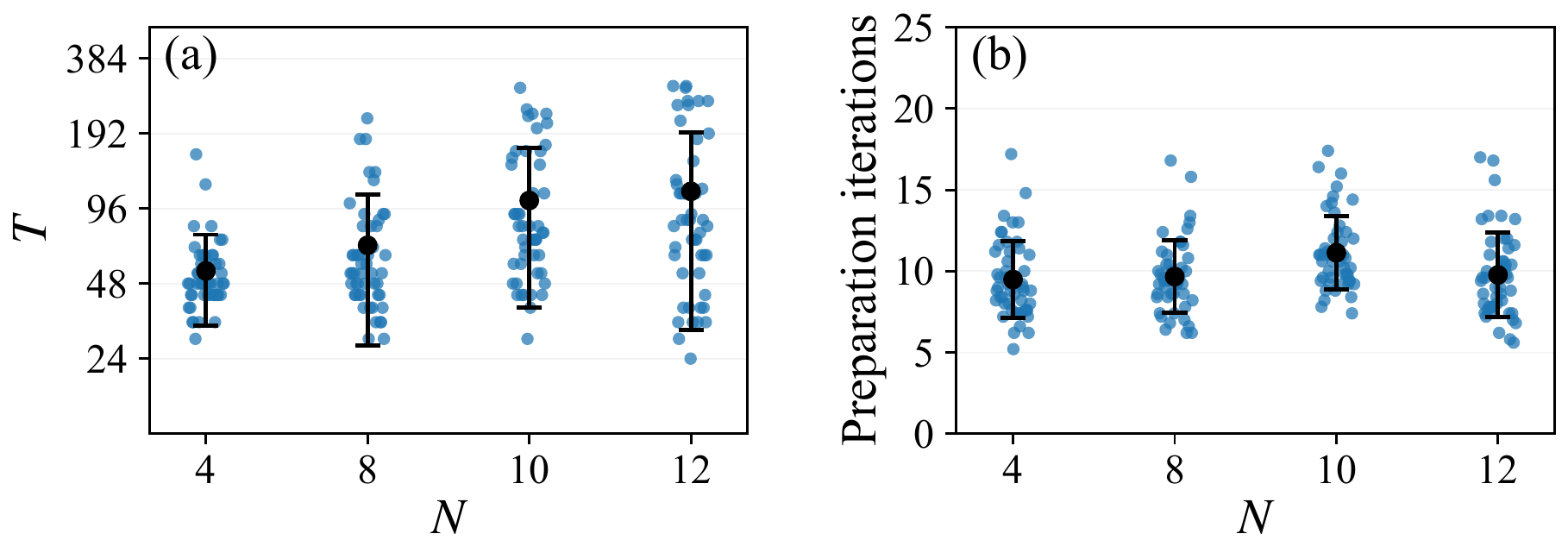}
\caption{(a) Rotation counts $T$ and (b) preparation iterations at each $T$ for sampled-\(\AHS\) preparation with the
same circuit and initialization as Fig.~\ref{fig:spin-exact-ahs-trajectories}.
For each initialization, the rotation count needed to reach $\AHS<10^{-2}$, evaluated using the exact SU(2) average, is determined by increasing $T$ through $24\times 2^k$, $k=0,\cdots,5$. 
Each blue dot shows the average of this count $T$ over five independent rotation samples in (a), and the corresponding average number of preparation sweeps in (b). One iteration corresponds to a sweep over the preparation parameters.
Black circles and error bars show the mean and standard deviations of these 50 averages. 
}
\label{fig:spin-sampled-rotation-requirement}
\end{figure}

The results in Fig.~\ref{fig:spin-exact-ahs-trajectories} use the exact SU(2) average for preparation. To examine whether the same asymmetry target can be reached with a finite sample of collective rotations, we replace this average by the sampled objective in Eq.~\eqref{eq:sampled-ahs}. The preparation circuit and initial parameter vectors are unchanged.
For each of the initial parameter vectors at each size, preparation is repeated with five independently sampled sets of Haar-random SU(2) rotations. Each set initially contains $T=24$ rotations and the number of rotations is doubled if the prepared state does not satisfy $\AHS<10^{-2}$, as evaluated using the exact SU(2) average. All combinations at each size reach this threshold.
Figure~\ref{fig:spin-sampled-rotation-requirement}a shows the smallest rotation count satisfying the criterion, $T$, averaged over the five rotation samples for each initialization. Panel (b) shows the preparation iterations needed to reach the same criterion at $T$ averaged over the same five samples.
Together, these results show that preparation with sampled rotations reaches the prescribed exact asymmetry threshold within a small number of preparation iterations across the tested system sizes and initializations.

\section{Discussion and conclusions}

We have developed symmetry-guided variational algorithms that use Hilbert--Schmidt asymmetry as an objective for state preparation and for parameter updates between energy minimization stages. The symmetry information enters through the group action, without requiring the preparation circuit to preserve the target sector. 
This step requires neither a classically computed approximation to the target wave function nor Hamiltonian energy evaluations.
Its integration with VQE provides warm start and alternating optimization procedures for using this information before and during the energy search.

The nuclear and spin applications examine this approach in spaces with different degrees of freedom and singlet multiplicities. In the $p+{}^3\mathrm{He}$ model, asymmetry guidance improves the reliability of reaching joint energy and symmetry targets while the spin and radial parameters are optimized with a generic ansatz. 
In the spin models, preparation consistently increases the singlet probability across different initializations and system sizes, and subsequent exchange optimization preserves this probability while lowering the energy. The combined procedure yields higher ground-state fidelities than VQE for the larger instances. At $N=12$, this improvement occurs even though preparation reduces the initial ground-state overlap. 
The benefit of our $\AHS$ preparation therefore need not come from approximating the target wave function more closely at the outset. The preparation can instead provide an input concentrated in the desired sector. 
These results illustrate a route to improving variational calculations through symmetry-guided optimization, distinct from adaptive ansatz construction \cite{Grimsley2019ADAPT}.

Preparation with finite rotation samples reaches the prescribed asymmetry threshold in the spin models. The purity and rotated-overlap terms that define this sampled objective can be estimated through measurements on two copies of the state, providing a way to directly evaluate the objective on a quantum device. 
The nuclear finite-shot calculations also examine optimization in the presence of measurement uncertainty.

Because the preparation objective is defined by the group action rather than the Hamiltonian energy, $\AHS$-based preparation may also provide inputs for other quantum algorithms requiring the corresponding symmetry. 
The same formulation motivates extensions to higher-rank symmetries, with potential applications to nuclear model spaces organized by SU(3) and U(3) \cite{Elliott1958,Rowe1985,LauneyPPNP2016}. In such settings, defining an objective through the group action offers a route that does not require explicitly constructing and measuring a set of Casimir operators.
Extensions would require identifying suitable group representations and establishing how the corresponding asymmetry objective relates to the desired target subspace.

\ack{
We thank Alexis Mercenne and Mark M. Wilde for helpful discussions. This work was supported by the U.S. Department of Energy (DOE), DE-SC0023694, This work was performed in part at the Aspen Center for Physics, which is supported by National Science Foundation grant PHY-2210452.
This research benefited from resources of the Oak Ridge Leadership Computing Facility at the Oak Ridge National Laboratory, which is supported by the Office of Science of the U.S. Department of Energy under Contract No. DE-AC05-00OR22725.
The work benefited from high performance computational resources provided by LSU (www.hpc.lsu.edu), the National Energy Research Scientific Computing Center (NERSC), a U.S. Department of Energy Office of Science User Facility at Lawrence Berkeley National Laboratory operated under Contract No. DE-AC02-05CH11231, as well as the Frontera computing project at the Texas Advanced Computing Center, made possible by National Science Foundation award OAC-1818253. 
}



\data{The data that support the findings of this study are available from the corresponding author upon reasonable request.}



\appendix

\section{Circuits for asymmetry estimation and variational ansatzes}
\label{app:circuits_ansatzes}

\begin{figure}
\centering
\includegraphics[width=0.85\textwidth]{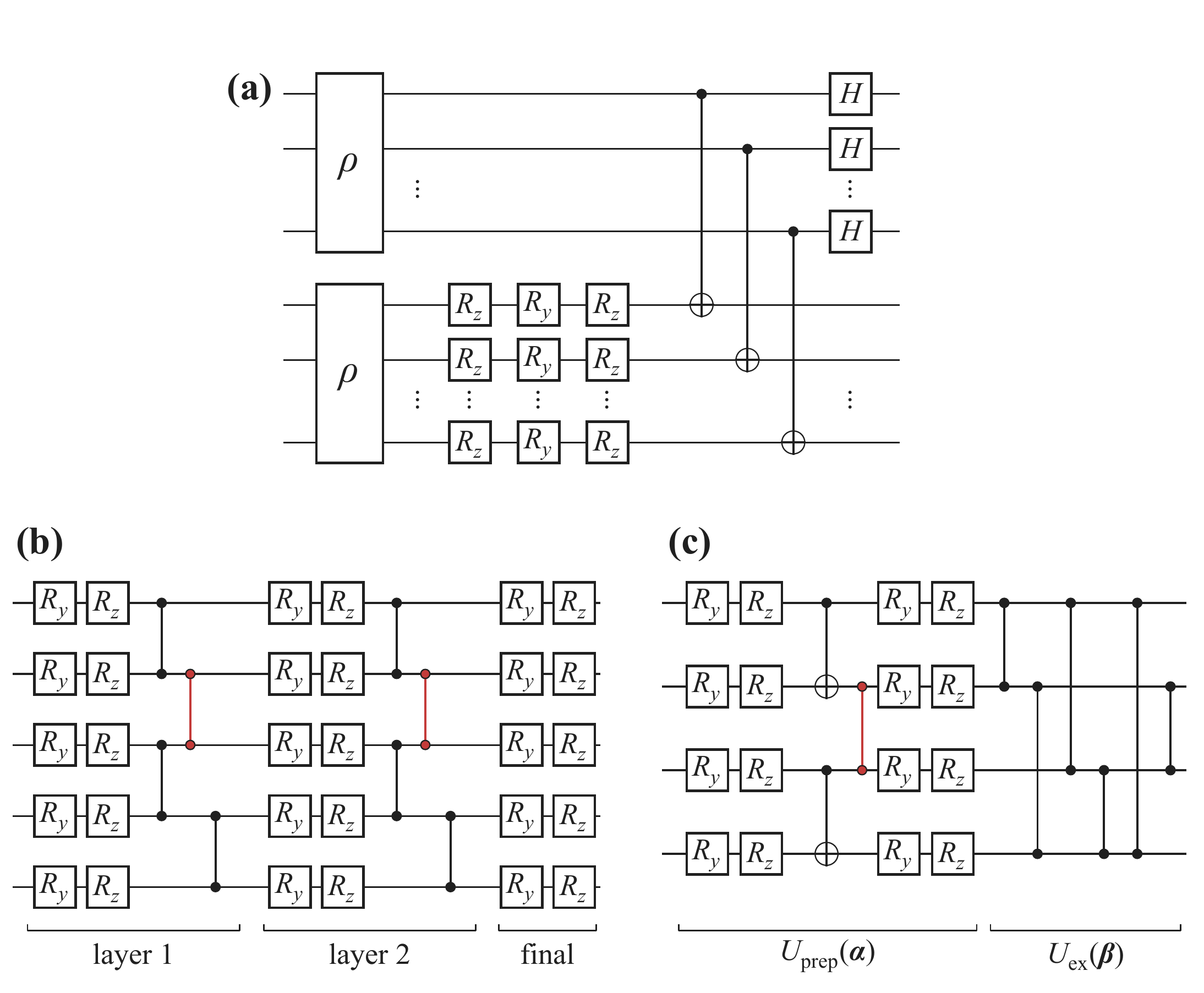}
\caption{Circuits for asymmetry estimation and quantum variational ansatzes. 
(a) Rotated overlap circuit for two copies of $\rho$, implemented in Ref.~\cite{2023Bandyopadhyay}. The rotation gates act from left to right as $R_z(\gamma)$, $R_y(\beta)$ and $R_z(\alpha)$ with the same Euler angles on every qubit of the lower copy. 
(b) Nuclear variational ansatz with five qubits and two layers. The upper two qubits encode the cluster spins, and the lower three encode the radial basis states. Each layer applies $R_y$ followed by $R_z$ to every qubit and then $R_{xx}$ gates on the indicated pairs. A final set of $R_y$ and $R_z$ rotations completes the circuit.
(c) Neutrino spin variational ansatz with four spins as an example. The preparation circuit applies local $R_y$ and $R_z$ rotations, CX gates on disjoint pairs, and final local rotations. It is followed by one layer of exchange gates over all spin pairs. 
Connected filled circles denote $R_{xx}$ in (b) and the exchange gates of Eq.~\eqref{eq:spin-exchange-gate} in $U_{\mathrm{ex}}$ in (c). Red gates are the additional $R_{xx}$ gates coupling the spin and radial qubits in (b), and the $R_{zz}$ gate coupling the preparation pairs in (c), used in Appendices~\ref{app:nuclear-entanglement} and~\ref{app:spin-additional-entanglers}, respectively. 
}
\label{fig:nuclear_and_spin_ansatzes}
\end{figure}

Figure~\ref{fig:nuclear_and_spin_ansatzes}a shows the circuit for estimating the rotated overlaps $\operatorname{Tr}[\rho U(g)\rho U^\dagger(g)]$ in Eq.~\eqref{eq:sampled-ahs} using the destructive SWAP test of Ref.~\cite{2023Bandyopadhyay}. According to Algorithm 2 of Ref.~\cite{2023Bandyopadhyay}, two copies of $\rho$ are prepared, and a collective rotation is applied to the lower copy, followed by CX gates between corresponding qubits and $H$ gates on the upper qubits. The overlap is then estimated from computational basis measurement outcomes. Omitting the rotations gives $\operatorname{Tr}(\rho^2)$, the purity term in the same equation. Repeating the overlap estimate for the sampled rotations provides the group average used to evaluate $\AHS^{\mathrm{sampled}}$ in Eq.~\eqref{eq:sampled-ahs}.

Panels (b) and (c) relate this measurement circuit to the variational states used in the two applications.
In the nuclear model, the state $\rho$ in panel (a) is prepared by the separate two qubit spin circuit shown in panel (b). The energy is evaluated as the Hamiltonian expectation value in the full five qubit state prepared by the circuit in panel (b). 
In the spin model, $\rho$ is prepared by $U_{\mathrm{prep}}(\bm\alpha)$ in panel (c) acting on $|\Phi_p\rangle$ in Eq.~\eqref{eq:spin-trial}. The subsequent circuit $U_{\mathrm{ex}}(\bm\beta)$ applies the exchange gates defined in Eq.~\eqref{eq:spin-exchange-gate} and preserves the prepared total spin sector probabilities. 
Red gates allow the additional entanglement between the spin and radial degrees of freedom in panel (b), and between the preparation pairs in panel (c). The corresponding comparisons are presented in Appendices~\ref{app:nuclear-entanglement} and~\ref{app:spin-additional-entanglers}, respectively.

\section{Conditional objective for spin optimization}
\label{app:conditional-objective}

While the $\AHS$ measure is entirely determined on the quantum processor according to Fig.~\ref{fig:nuclear_and_spin_ansatzes}, in this section we provide the reasoning behind our choice for the asymmetry cost function of Eq.~\eqref{eq:angular-stage-objective}.
This objective uses the sampled asymmetry $\AHS^{\mathrm{sampled}}$ to determine whether each trial state is evaluated through its energy or an asymmetry-dependent cost. Above the empirical threshold $\tau$, the cost is chosen to favor lower asymmetry. At or below $\tau$, the objective is the energy at the current radial parameters.
For a pure two-qubit spin state, $\rho=|\psi_{\mathrm{spin}}\rangle \langle\psi_{\mathrm{spin}}|$, Eq.~\eqref{eq:sampled-ahs} gives
\begin{equation}
1 - \frac{\AHS^{\mathrm{sampled}}}{2}
= \frac{1}{T} \sum_{t=1}^T |\langle\psi_{\mathrm{spin}}|U(g_t)|\psi_{\mathrm{spin}}\rangle|^2.
\end{equation}
The chosen cost is inversely proportional to the square root of this average overlap with $|E_{\mathrm{ref}}|$ setting the scale. For fixed, nonzero $E_{\mathrm{ref}}$, it therefore increases with $\AHS^{\mathrm{sampled}}$ for $\tau < \AHS^{\mathrm{sampled}} < 2$. Minimizing this cost favors larger overlaps between the state and its sampled rotations without evaluating the energy of each trial state above the threshold.

\section{Nuclear calculations with spin-radial entanglement}
\label{app:nuclear-entanglement}

\begin{figure}[htbp]
\centering
\includegraphics[width=0.75\textwidth]{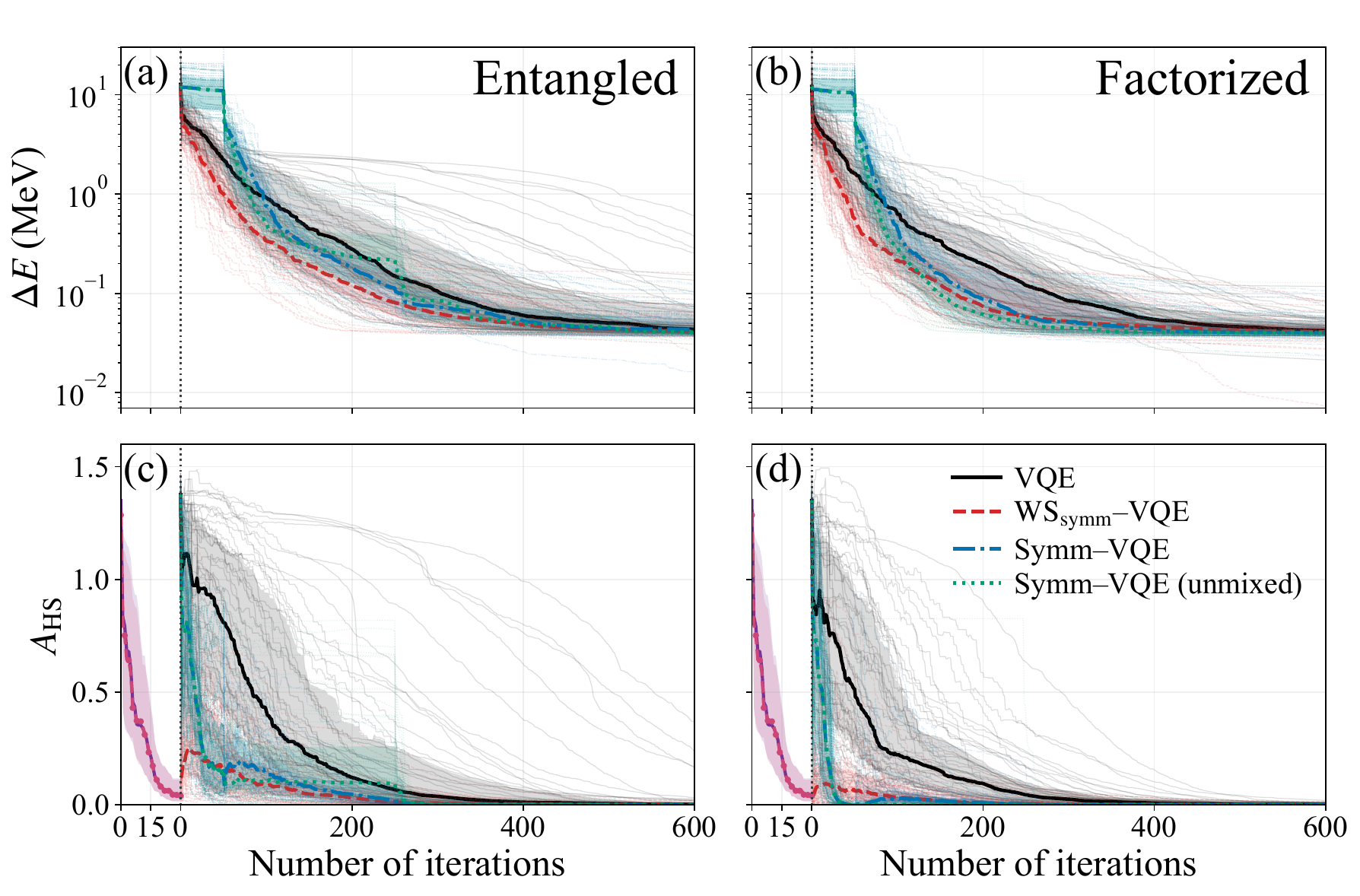}
\caption{Convergence of the three $\AHS$-guided algorithms and VQE for the Entangled (left) and Factorized (right) circuits.  
Panels (a) and (b) show the energy error $\Delta E$, and panels (c) and (d) show the asymmetry $\AHS$. 
Thin curves show individual results for 50 initializations, while thick curves and shaded bands show medians and interquartile ranges. The optimizer and initialization settings are the same as in Fig.~\ref{fig:noiseless-iteration-trajectories}. 
WS preparation is shown to the left of the vertical dotted line in (c) and (d), with both exact and sampled asymmetry.
}
\label{fig:nuclear-individual-trajectories}
\end{figure}

A parametrized $R_{xx}$ gate on $(q_1, q_2)$, shown in red in Fig.~\ref{fig:nuclear_and_spin_ansatzes}b, is added to each variational layer, with the additional angles denoted by $\bm\theta_{mr}$, 
allowing entanglement between the spin and radial degrees of freedom. All other gates and the reference state are unchanged. The resulting circuit has 38 parameters and is labeled ``Entangled" in Fig.~\ref{fig:nuclear-individual-trajectories}, while the circuit used in the main text is labeled ``Factorized".
The additional parameters $\bm\theta_{mr}$ are held fixed during $\mathrm{WS}_{\mathrm{symm}}$--VQE preparation and the alternating spin stages, and varied during energy minimization. 
Symm-VQE (unmixed) optimizes both $\bm\theta_r$ and $\bm\theta_{mr}$, extending the radial update of the main text to include the additional $R_{xx}$ gate angles. The remaining optimization settings are unchanged. 
For the Entangled circuit, the full state asymmetry at the start of energy optimization differs from the spin asymmetry reached during $\mathrm{WS}_{\mathrm{symm}}$--VQE preparation, since the preparation objective is evaluated on the separate two qubit spin circuit.
In the preparation interval of Fig.~\ref{fig:nuclear-individual-trajectories}, purple solid and magenta dashed curves show the exact and sampled asymmetry, respectively.

The Factorized panels in Fig.~\ref{fig:nuclear-individual-trajectories} show the same outcomes as in Fig.~\ref{fig:noiseless-iteration-trajectories}, but showing all individual
50 initializations
(see the thin curves in Fig.~\ref{fig:nuclear-individual-trajectories}).
For the Entangled circuit, the three $\AHS$-guided algorithms show lower median energy errors than VQE during optimization, as in the Factorized case. All four algorithms achieve median final fidelities above $99.5\%$.

\section{Finite-shot measurements for the nuclear model}
\label{app:finite-shot-nuclear-raw-data}

\begin{figure}[htbp]
\centering
\includegraphics[width=0.85\textwidth]{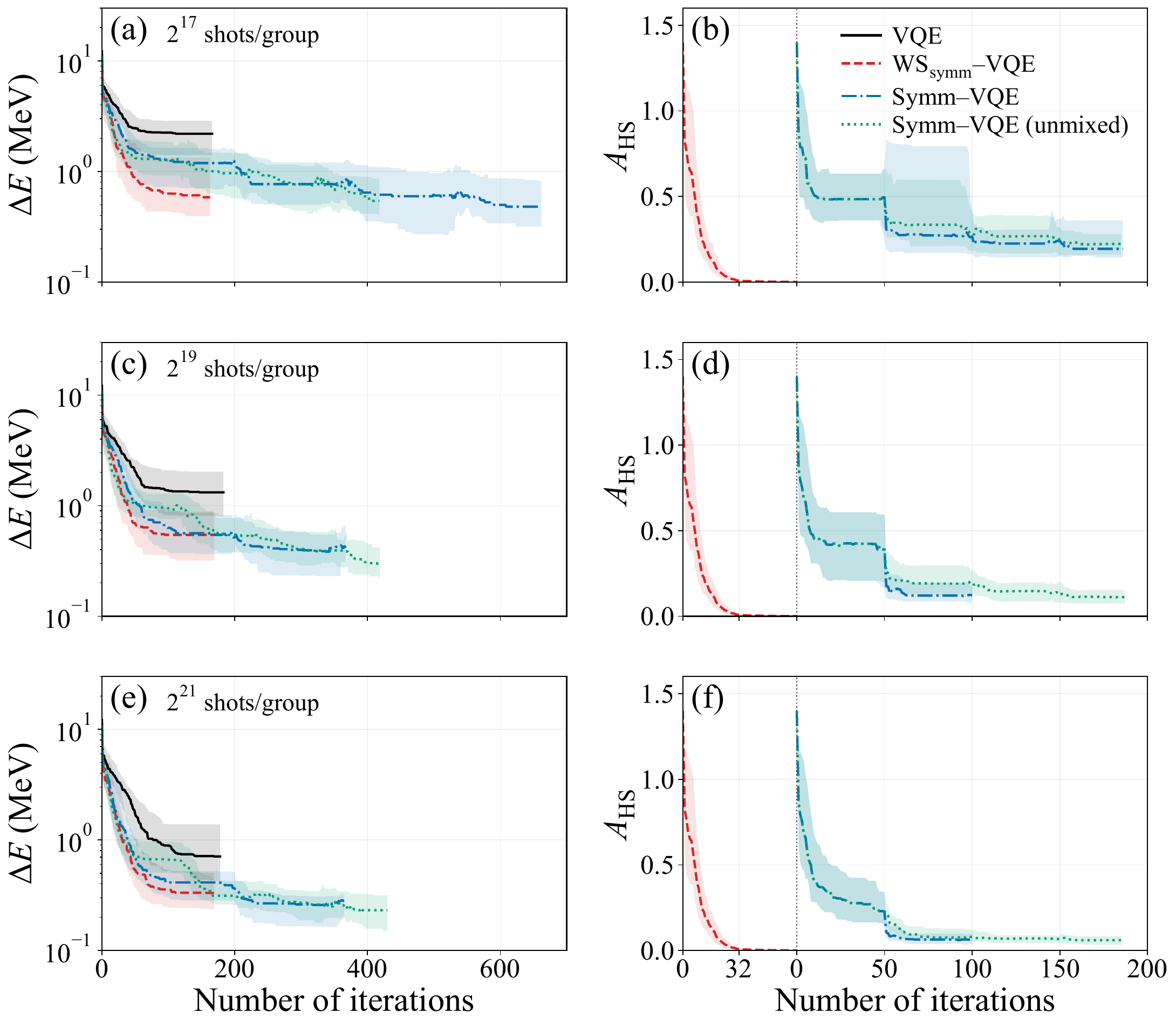}
\caption{
Measured energy error (left) and sampled asymmetry (right) for the four algorithms under finite-shot optimization. Rows correspond to $2^{17},2^{19}$, and $2^{21}$ energy shots per qubit-wise commuting group. Colors and line styles are the same as in Fig.~\ref{fig:nuclear-finite-shots-SV-analysis}. Curves and shaded bands show medians and interquartile ranges over the same 50 initial parameter vectors. Preparation for ${\mathrm{WS}_{\mathrm{symm}}}$--VQE is shown to the left of the vertical dotted line in the right panels.
}
\label{fig:nuclear-finite-shots-raw-data}
\end{figure}

Figure~\ref{fig:nuclear-finite-shots-raw-data} shows the energy error and sampled asymmetry measured during the finite-shot optimization of Sec.~\ref{sec:nuclear-simulation-results},
while Fig.~\ref{fig:nuclear-finite-shots-SV-analysis} shows statevector evaluations at the same parameters. 
The left panels show energy minimization, with iterations accumulated across the energy stages of the alternating algorithms and the intervening spin stages omitted. The right panels likewise accumulate the spin iterations across cycles, with the energy stages omitted.

\section{Spin model with additional entangling gates}
\label{app:spin-additional-entanglers}

\begin{figure}[htbp]
\centering
\includegraphics[width=.8\textwidth,height=.66\textheight,keepaspectratio]{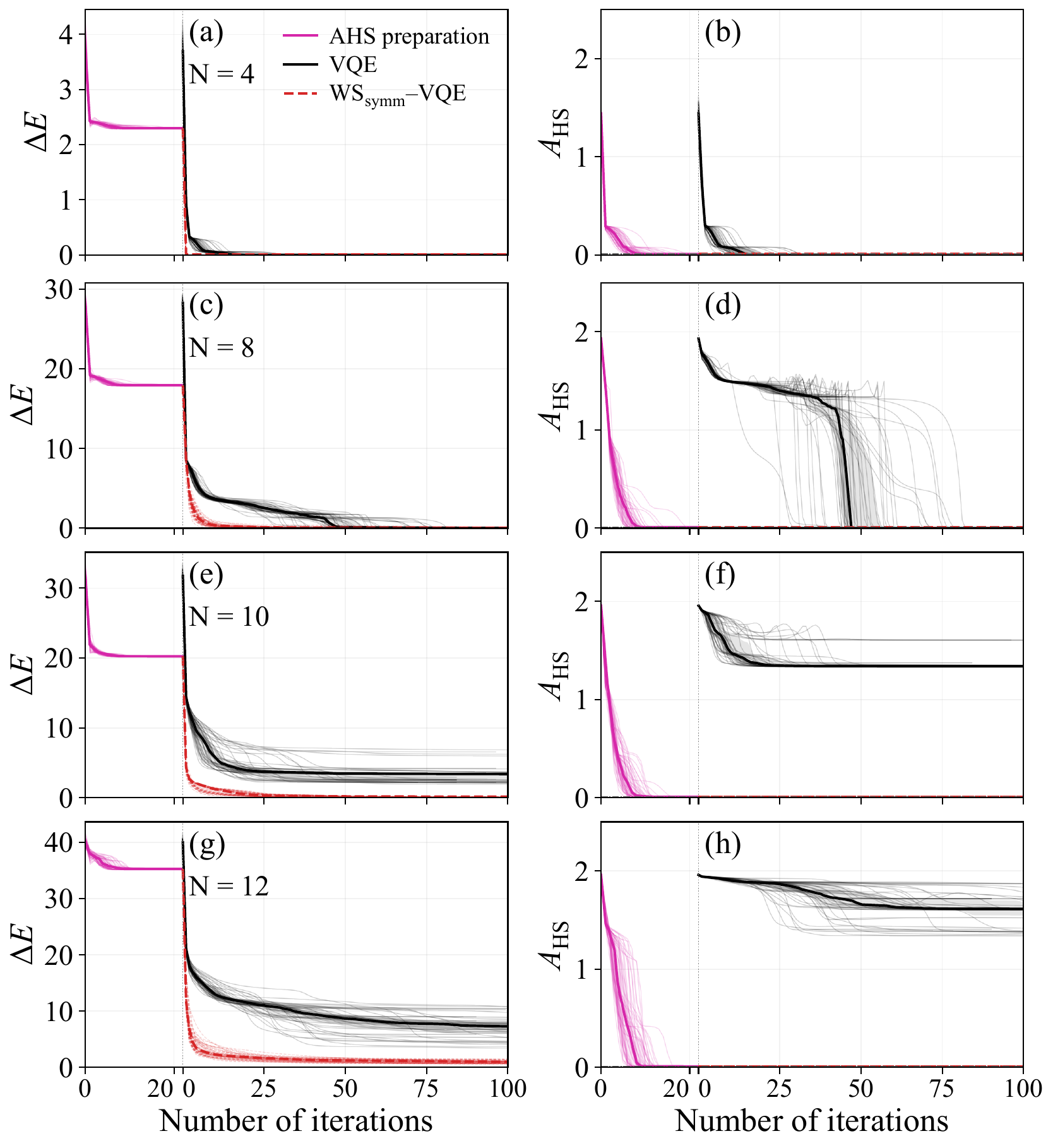}
\caption{Energy error (left) and asymmetry $\AHS$ (right) for $N=4,8,10,12$ with additional inter-pair $R_{zz}$ gates in the preparation circuit. 
Magenta denotes $\AHS$ preparation, black VQE and red dashed ${\mathrm{WS}_{\mathrm{symm}}}$--VQE. Thin curves show 50 initializations, and thick curves and bands give medians and interquartile ranges with terminal values carried forward.  
One iteration corresponds to a complete coordinate sweep over the parameters optimized in each stage. The optimizer and settings are the same as in Fig.~\ref{fig:spin-exact-ahs-trajectories}.
}
\label{fig:spin-additional-entanglers}
\end{figure}

The preparation circuit in Sec.~\ref{sec:spin-models} entangles disjoint pairs of spins (particles). To allow entanglement between these pairs, we insert parametrized $R_{zz}$ gates on $(1,2),(3,4),\cdots,(N-3,N-2)$ between the CX gates and the final local rotation layer. 
The additional $N/2-1$ angles are appended to the $4N$ local rotation parameters in $\bm\alpha$ and initialized from $0.1\mathcal{N}(0,1)$. The exchange parameters $\bm\beta$ are unchanged.
The comparison uses the same 50 initializations at each size, with all other circuit and optimization settings unchanged from the main text.

Figure~\ref{fig:spin-additional-entanglers} shows that all preparation runs reach $\AHS<10^{-2}$. With the additional entangling gates, the median final fidelity of VQE increases from 29.63 $\%$ to 47.65 $\%$ at $N=10$ and from $0.02\%$ to 26.23 $\%$ at $N=12$. The corresponding ${\mathrm{WS}_{\mathrm{symm}}}$--VQE fidelities remain close to those in the main text, changing from 99.10 $\%$ to 99.12 $\%$ and from 95.27 $\%$ to 94.94 $\%$, respectively.
Although the additional entanglers improve direct VQE, it still fails to converge to the ground state for most initializations at $N=10$ and 12.

\section{$N=6$ spin cases with $J=1$}
\label{app:N6_spin_cases}

\begin{figure}[htbp]
\centering
\includegraphics[width=.85\textwidth,height=.76\textheight,keepaspectratio]{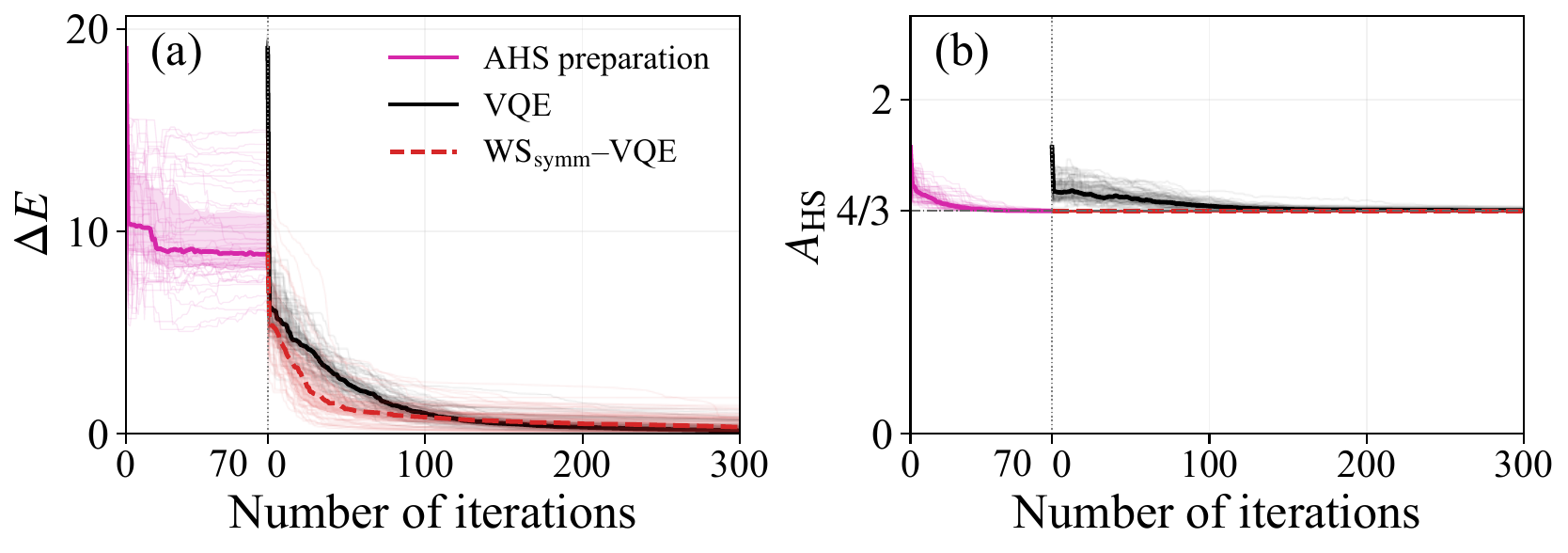}
\caption{
Energy Error (a) and asymmetry (b) for $N=6$ with a $J=1$ ground state using COBYLA. The preparation circuit preserves $M=1$. Magenta and red dashed curves show asymmetry preparation and subsequent exchange optimization, respectively. Thin curves show individual runs and thick curves and shaded bands indicate the median and interquartile range over 50 initializations. The vertical line separates the stages.
}
\label{fig:N6_spin_case}
\end{figure}

To illustrate asymmetry-guided optimization beyond singlet targets, we consider
the $N=6$ instance of Eq.~\eqref{eq:neutrino-hamiltonian}, whose ground state has $J=1$. To target this sector, we use a preparation circuit that preserves $M=1$, restricting the accessible sectors to $J=1,2,3$.
Figure~\ref{fig:N6_spin_case} shows the energy error and asymmetry for VQE and $\mathrm{WS}_{\mathrm{symm}}-\mathrm{VQE}$.
Starting from $|001100\rangle$, the preparation circuit applies two layers of local $R_z$ rotations and gates $\exp[-i\alpha(X_iX_j+Y_iY_j)/2]$, followed by final local $R_z$ rotations. The two-qubit gates act on $(0,1),(2,3),(4,5),(1,2),(3,4)$ in each layer, giving $28$ preparation parameters. The exchange circuit is unchanged from Eq.~\eqref{eq:spin-exchange-gate}.
For a pure state at fixed $M=1$, Eq.~\eqref{eq:ahs-general} reduces to
\begin{equation}
\AHS=2\left(1-\frac{P_1^2}{3} -\frac{P_2^2}{5} -\frac{P_3^2}{7}\right), 
\end{equation}
where $P_J$ is the probability of total spin $J$ and $P_1+P_2+P_3=1$. 
The minimum, $\AHS=4/3$, is attained at $P_1=1$. Minimizing asymmetry therefore selects $J=1$ within this fixed-$M$ space.

\bibliographystyle{iopart-num}
\bibliography{main_references}

\end{document}